\documentclass[traditabstract]{aa} % for the abstract without structuration % (traditional abstract) 
\usepackage{graphicx}
\usepackage{txfonts}
\usepackage{natbib}
\usepackage{longtable}
\def\ms{\hbox{\,m\,s$^{-1}$}}         %m.s -1
\def\m2s2{\hbox{\,m$^{2}$\,s$^{-2}$}} %m2.s -2
\def\kms{\hbox{\,km\,s$^{-1}$}}       %km.s -1
\def\Msun{\hbox{$\mathrm{M}_{\odot}$}}             %Msun
\def\Rsun{\hbox{$\mathrm{R}_{\odot}$}}
\def\Mjup{\hbox{$\mathrm{M}_{\rm Jup}$}}

\def\Me{\hbox{$\mathrm{M}_{\oplus}$}}             %Msun
\def\Re{\hbox{$\mathrm{R}_{\oplus}$}}

\def\mp{M_{\rm p}}
\def\rp{R_{\rm p}}

\begin{document}

\title{Characterization of the Kepler-101 planetary system with HARPS-N\thanks{Based on observations made with the Italian Telescopio Nazionale Galileo (TNG) operated on the island of La Palma by the Fundaci\'on Galileo Galilei of the INAF (Istituto Nazionale di Astrofisica) at the Spanish Observatorio del Roque de los Muchachos of the Instituto de Astrofisica de Canarias.} }
\subtitle{A hot super-Neptune with an Earth-sized low-mass companion}
\titlerunning{Characterization of the Kepler-101 planetary system with HARPS-N}
\authorrunning{Bonomo et al. 2014}

\author{   A.~S.~Bonomo\inst{1}  
\and		A.~Sozzetti\inst{1} 
\and		C.~Lovis\inst{2}   		 
\and		L.~Malavolta\inst{3,  4} 
\and		K.~Rice\inst{5}  
\and		L.~A.~Buchhave\inst{6, 7} 
\and		D.~Sasselov\inst{6}  
\and		A.~C.~Cameron\inst{8}  
\and		D.~W.~Latham\inst{6} 		
\and		E.~Molinari\inst{9, 10} 
\and		F.~Pepe\inst{2} 
\and		S.~Udry\inst{2} 	
\and		L.~Affer\inst{11} 					
\and		D.~Charbonneau\inst{6}  
\and		R.~Cosentino\inst{9}  
\and		C.~D.~Dressing\inst{6}  
\and		X.~Dumusque\inst{6}
\and		P.~Figueira\inst{12}  
\and		A.~F.~M.~Fiorenzano\inst{9}  
\and		S.~Gettel\inst{6}  
\and		A.~Harutyunyan\inst{9}  
\and		R.~D.~Haywood\inst{8}  
\and		K.~Horne\inst{8}  
\and		M.~Lopez-Morales\inst{6}  
\and		M.~Mayor\inst{2}  
\and		G.~Micela\inst{11}  
\and		F.~Motalebi\inst{2}  
\and		V.~Nascimbeni\inst{3}  
\and		D.~F.~Phillips\inst{6}  
\and		G.~Piotto\inst{3, 4}  
\and		D.~Pollacco\inst{13}  
\and		D.~Queloz\inst{2, 14}  
\and		D.~S\'egransan\inst{2} 
\and		A.~Szentgyorgyi\inst{6}  
\and		C.~Watson\inst{15}
		}

\institute{
INAF - Osservatorio Astrofisico di Torino, via Osservatorio 20, 10025 Pino Torinese, Italy  
\and Observatoire Astronomique de l'Universit\'e de Gen\`eve, 51 ch. des Maillettes, 1290 Versoix, Switzerland  
\and Dipartimento di Fisica e Astronomia ``Galileo Galilei", Universit\`a di Padova, Vicolo dell'Osservatorio 3, 35122 Padova, Italy  
\and INAF - Osservatorio Astronomico di Padova, Vicolo dell'Osservatorio 5, 35122 Padova, Italy  
\and SUPA, Institute for Astronomy, Royal Observatory, University of Edinburgh, Blackford Hill, Edinburgh EH93HJ, UK  
\and Harvard-Smithsonian Center for Astrophysics, 60 Garden Street, Cambridge, Massachusetts 02138, USA  
\and Centre for Star and Planet Formation, Natural History Museum of Denmark, University of Copenhagen, DK-1350 Copenhagen, Denmark 
\and SUPA, School of Physics \& Astronomy, University of St. Andrews, North Haugh, St. Andrews Fife, KY16 9SS, UK  
\and INAF - Fundaci\'on Galileo Galilei, Rambla JosŽ Ana Fernandez PŽrez 7, 38712 Bre–a Baja, Spain  
\and INAF - IASF Milano, via Bassini 15, 20133, Milano, Italy  
\and INAF - Osservatorio Astronomico di Palermo, Piazza del Parlamento 1, 90124 Palermo, Italy  
 \and Centro de Astrof\`isica, Universidade do Porto, Rua das Estrelas, 4150-762 Porto, Portugal  
\and Department of Physics, University of Warwick, Gibbet Hill Road, Coventry CV4 7AL, UK  
\and Cavendish Laboratory, J J Thomson Avenue, Cambridge CB3 0HE, UK  
\and Astrophysics Research Centre, School of Mathematics and Physics, Queens University, Belfast, UK }

\date{Received... / Accepted ...}

\offprints{\\
\email{bonomo@oato.inaf.it}}

\abstract{We report on the characterization of the Kepler-101 planetary system, thanks to 
a combined DE-MCMC analysis of \emph{Kepler} data and forty 
radial velocities obtained with the HARPS-N spectrograph. 
This system was previously validated by Rowe et al. (2014) and is composed of a hot super-Neptune, Kepler-101b, 
and an Earth-sized planet, Kepler-101c. These two planets orbit the slightly evolved and metal-rich G-type star in 3.49 and 6.03~days, respectively. 
With mass $\mp=51.1_{-4.7}^{+5.1}~\Me$, radius $\rp=5.77_{-0.79}^{+0.85}~\Re$,
and density $\rho_{\rm p}=1.45_{-0.48}^{+0.83}~\rm g\;cm^{-3}$, 
Kepler-101b is the first fully-characterized super-Neptune, and its density suggests that 
heavy elements make up a significant fraction of its interior; more than 60\% of its total mass.
Kepler-101c has a radius of $1.25_{-0.17}^{+0.19}~\Re$, which implies the absence of any H/He envelope, but
its mass could not be determined due to the relative faintness of the parent star  
for highly precise radial-velocity measurements ($K_{\rm p}=13.8$) and the limited number of radial velocities.
The $1~\sigma$ upper limit, $\mp < 3.8~\Me$, excludes a pure iron composition with a $68.3\%$ probability.
The architecture of the Kepler-101 planetary system - containing a close-in giant planet 
and an outer Earth-sized planet with a period ratio slightly larger than the 3:2 resonance - is certainly
of interest for planet formation and evolution scenarios. This system does not 
follow the trend, seen by Ciardi et al. (2013), that in the majority of \emph{Kepler} systems of planet pairs with at least 
one Neptune-size or larger planet, the larger planet has the longer period.
}

% 5 {} token are mandatory
\maketitle
%
%________________________________________________________________

 \section{Introduction}
Studies of planetary population synthesis within the context of the core accretion model (e.g., \citealt{IdaLin2008}; \citealt{Mordasinietal2009}) 
suggest that the planetary Initial Mass Function is characterized by physically significant minima and maxima. 
In particular, a minimum in the approximate range $30\lesssim M_p\lesssim70$ M$_\oplus$ 
is understood as evidence for a dividing line between planets dominated 
in their interior composition by heavy elements, and giant gaseous planets that undergo runaway gas accretion. 
Recent theoretical work \citep{Mordasinietal2012} has also reproduced the basic shape of the planetary mass - radius relation 
and its time evolution in terms of the fraction of heavy elements $Z = M_Z/M$ in a planet. In particular, the radius distribution 
is predicted to be bimodal, with a wide local minimum in the approximate range $6\lesssim R_p\lesssim8$ R$_\oplus$, roughly coinciding with the 
minimum in the range of planetary masses indicated above. Furthermore, as regards the possible architectures of multiple-planet systems, 
considerable attention has been devoted to gauging the likelihood of the formation and survival of terrestrial-type planets interior and exterior to 
close-in higher-mass objects that have undergone Type I or II migration. 
In particular, recent investigations have not only shown that hot Earths might be 
found in systems in which disk material has been shepherded by a migrating giant \citep{Raymondetal2008}, 
or that water-rich terrestrial planets can still 
form in the habitable zones of systems containing a hot Jupiter \citep{FoggNelson2007}, 
but also that terrestrial planets could be found just outside 
the orbit of a hot Jupiter in configurations with a variable degree of dynamical interaction \citep{Ogiharaetal2013}.

The class of transiting exoplanets is uniquely suited to provide powerful constraints on the theoretical predictions of the formation, structural, and dynamical evolution history of planetary systems such as those listed above. 
For example, the observed $R_{\rm p}-M_{\rm p}$ diagram points towards a paucity of planets with properties intermediate between those of Neptune and Saturn. In particular, the mass bin between 30 and 60 M$_\oplus$ 
and the radius bin between 5 and 7 R$_\oplus$, are amongst the most under-populated, despite the fact that objects with such characteristics 
should be relatively easy to find in high-precision photometric and spectroscopic datasets. Furthermore, data from the Kepler mission indicates 
that, for multiple-planet architectures in which one object is approximately Neptune-sized or larger, the larger planet is most often the planet with the longer period \citep{Ciardietal2013}, and also that in general a lack of companion planets 
in hot-Jupiter systems is observed (e.g., \citealt{Lathametal2011, Steffenetal2012}).

In this work we combine Kepler photometry with high-precision radial-velocity measurements
of the Kepler-101 two-planet system gathered in the context 
of the GTO program of the HARPS-N Consortium \citep{Pepeetal2013, Dumusqueetal2014}. 
Kepler-101 was initially identified as Kepler Object of Interest 46 
(KOI-46). It was subsequently validated by \citet{Roweetal2014}, 
who derived orbital periods of 3.49 d and 6.03 d, and planetary radii of 5.87 R$_\oplus$ 
and 1.33 R$_\oplus$ for Kepler-101b and Kepler-101c, respectively, 
and also carried out successful dynamical stability tests. 
Our combined spectroscopic and photometric analysis allows us to derive 
a dynamical mass for Kepler-101-b and to place constraints on that of Kepler-101c. 
The much improved characterization of the Kepler-101 system permits us to identify 
the first fully-characterized super-Neptune planet, and to provide the first observational constraints 
on the architecture of multiple-planet systems 
with close-in low-mass giants and outer Earth-sized objects in orbits not far from resonance.

\section{Data}

\subsection{Kepler photometry}
%The hosting star Kepler-101 is faint ($K_{p}=13.8$) when compared with stars usually 
%observed in high-precision radial-velocity searches. 
%Its IDs, coordinates, and magnitudes are listed in Table~\ref{starplanet_param_table}. 
Kepler-101 (see IDs, coordinates, and magnitudes in Table~\ref{starplanet_param_table})
is a relatively faint target ($K_{p}=13.8$) for high-precision radial-velocity searches.
It was observed by \emph{Kepler} for almost four years, from quarter Q1 up to 
quarter Q17, with the long-cadence (LC) temporal sampling of 29.4~min, and for ten months,
from quarter Q4 up to quarter Q7, in short-cadence (SC) mode, i.e. one point every 58.8~s.
The medians of the errors of individual photometric measurements are 
155 and 846~ppm for LC and SC data, respectively.

The \emph{Kepler} light curve shows distinct transits of the 3.5~d transiting planet Kepler-101b, 
with a depth of $\sim 0.1\%$. On the other hand, the transits of the Earth-sized companion Kepler-101c,
with a period of 6.03~d and a depth of $\sim 55$~ppm, are embedded in the photon noise.
They can be detected after removing the Kepler-101b transits and by using data from more than 4-5 quarters,
because the phase-folded transit has a low S/N of $\sim 11$ 
when taking all the available LC measurements into account.

The simple-aperture-photometry\footnote{http://keplergo.arc.nasa.gov/PyKEprimerLCs.shtmlp} 
\citep{Jenkinsetal2010} measurements were used for the characterization of 
Kepler-101b and c (Sect.~\ref{sys_param}) and were corrected from flux contamination by background 
stars which are located in the \emph{Kepler} mask of our target. This amounts to only a few percent, 
as estimated by the \emph{Kepler} team\footnote{http://archive.stsci.edu/kepler/kepler\_fov/search.php}. 
  
No clear activity features with amplitude larger than $\sim~400$~ppm are seen in \emph{Kepler} LC data, 
indicating that the host star is magnetically quiet. %, and therefore old.

\subsection{Spectroscopic  follow-up with HARPS-N}
\subsubsection{Radial-velocity observations}
Forty spectra of Kepler-101, with exposure times of half an hour and average S/N of 16 at 550~nm, 
were obtained with the high-resolution ($R\sim115,000$) fiber-fed, optical echelle HARPS-N spectrograph, installed during Spring 2012 on the 3.57-m 
Telescopio Nazionale Galileo (TNG) at the Observatorio del Roque de los Muchachos, La Palma Island, Spain 
\citep{Cosentinoetal2012}.
HARPS-N is a near-twin of the HARPS instrument mounted at the ESO 3.6-m telescope in La Silla \citep{Mayoretal2003}
optimized for the measurement of high-precision radial velocities ($\lesssim1.2$ m s$^{-1}$ in 1 hr integration 
for a $V\sim12$ mag, slowly rotating late-G/K-type dwarf). Spectroscopic measurements of Kepler-101 were gathered 
with HARPS-N in obj\_AB observing mode, i.e. without acquiring a simultaneous Thorium
lamp spectrum. Indeed, for this faint target, the instrumental drift during one night is considerably
lower than the photon noise uncertainties. 
The first ten spectra were taken between June and August 2012, before the
readout of the red side of the CCD failed, which occurred in late September 2012. 
Thirty additional measurements were collected from 
the end of May 2013 up to the end of August 2013, 
after the replacement of the CCD.

\begin{figure*}[t]
\centering
\begin{minipage}{15cm}
%\vspace{0.5cm}
\includegraphics[width=7cm]{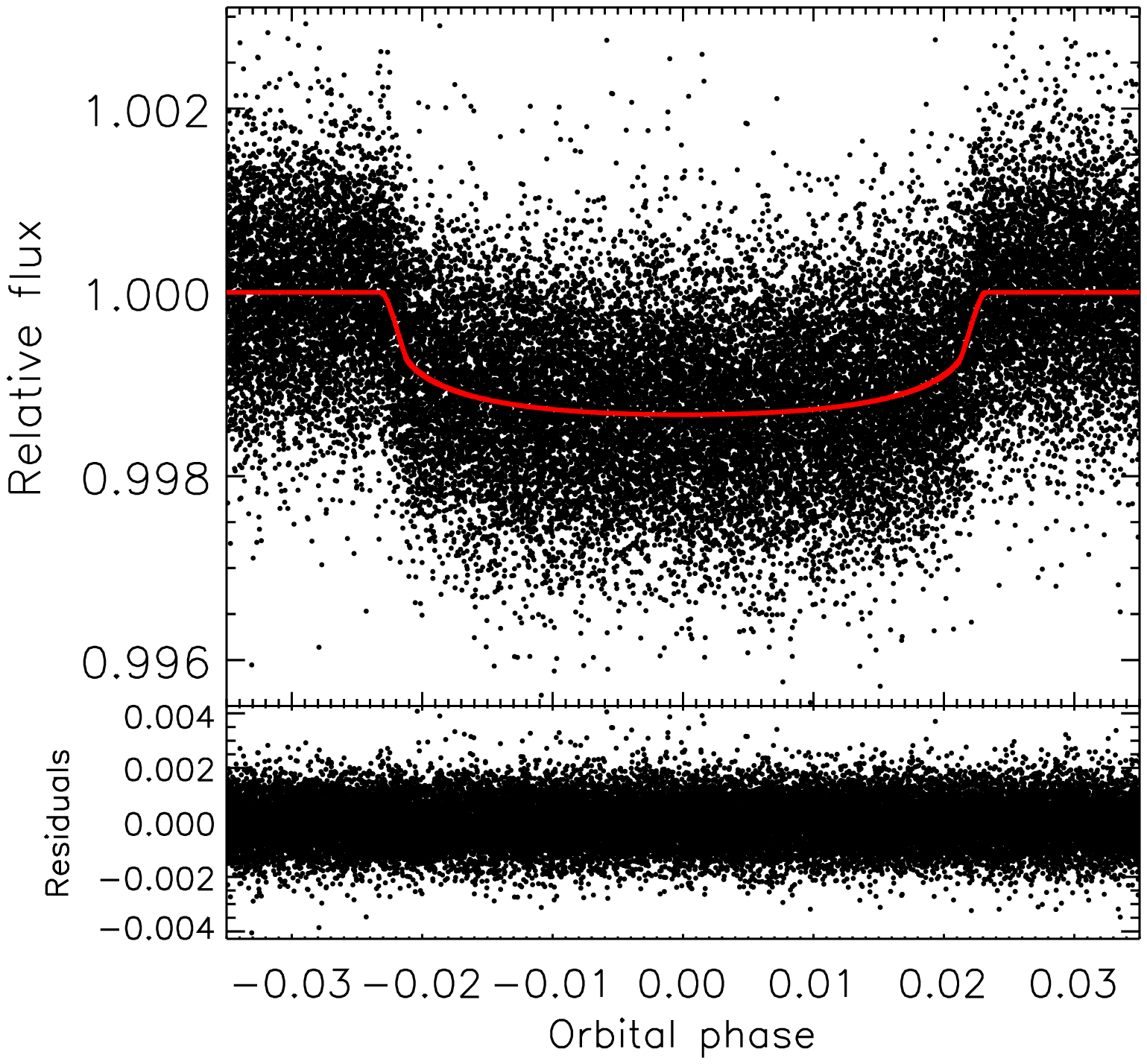}
\hspace{1.0cm}
\includegraphics[width=7cm]{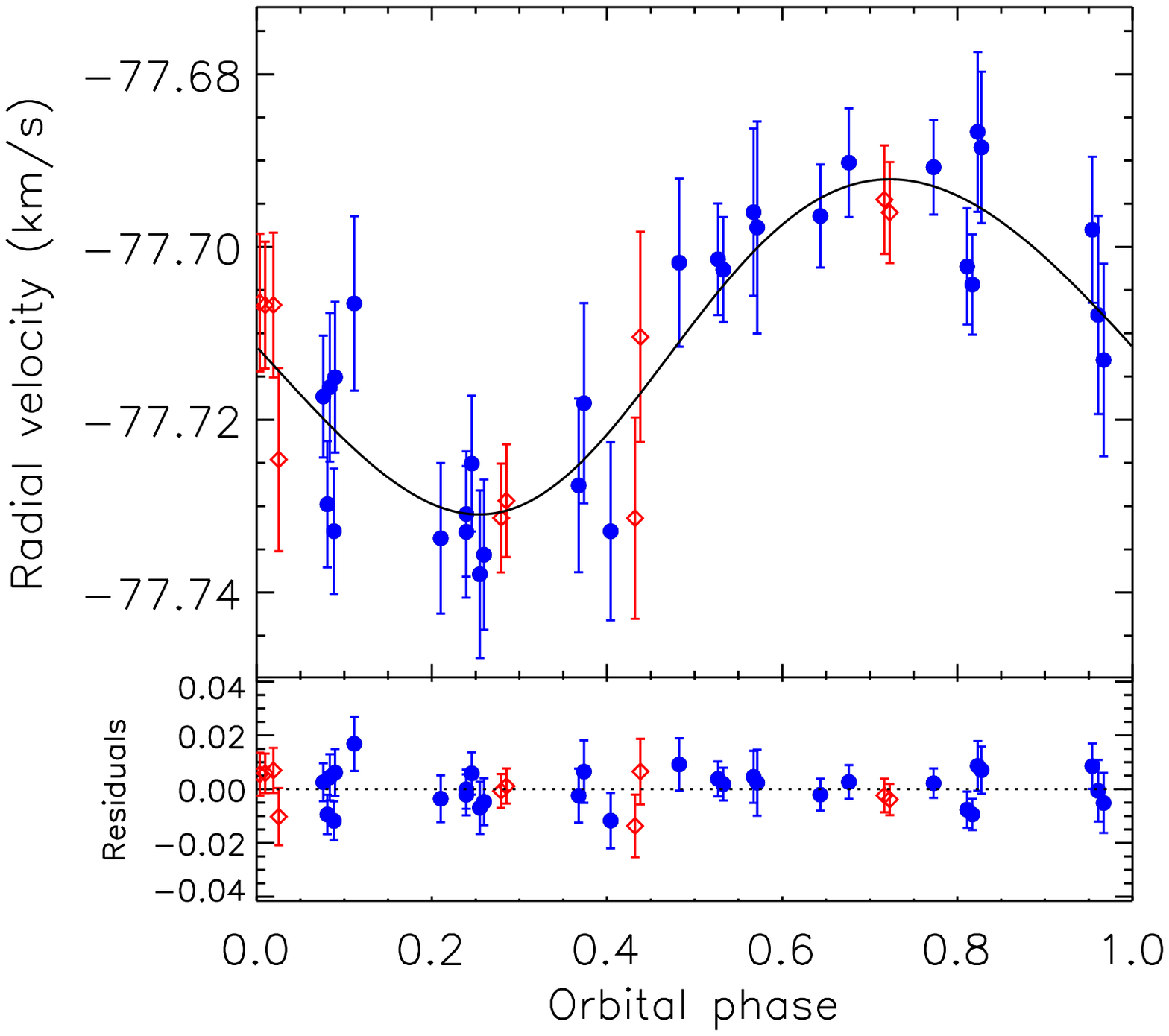}
\vspace{0.5cm}
\caption{
\emph{Left panel}: phase-folded short-cadence transit light curve of Kepler-101b along with the transit
model (red solid line).
\emph{Right panel}: phase-folded radial-velocity curve of Kepler-101b and, superimposed, the 
Keplerian orbit model (black solid line). Red and blue circles show the HARPS-N data obtained with
the original and replaced CCD, respectively. }
\label{fig_Kepler101b}
\end{minipage}
\end{figure*}

HARPS-N spectra were reduced with the on-line standard pipeline, and radial velocities 
were measured by means of a weighted cross correlation with a 
numerical spectral mask of a G2V star (\citealt{baranne96};  \citealt{pepe02}). 
They are listed in Table~\ref{table_rv} along with 
their $1~\sigma$ photon-noise uncertainties\footnote{RV jitter in our data is negligible as expected from the 
low magnetic activity level of Kepler-101.},
which range from 5.5 to 12.5~$\ms$, and bisector spans. 
HARPS-N radial velocities show a clear variation, 
with a semi-amplitude of $19.4 \pm 1.8 \ms$, 
in phase with the Kepler-101b ephemeris as derived from \emph{Kepler} photometry 
(see Fig.~\ref{fig_Kepler101b}). 
As expected from the relative faintness of the host star for high-precision radial velocities
and the limited number of HARPS-N measurements,
the RV signal induced by the Earth-sized planet Kepler-101c is not detected, 
hence only an upper limit can be placed on its mass. 

No correlation or anti-correlation between bisector spans and RVs is seen, 
as expected when RV variations are induced by planetary companions.

\subsubsection{Stellar atmospheric parameters}
\label{star_atmos_param}
We applied two slightly different approaches to derive the photospheric parameters of Kepler-101. 
The co-addition of all the available HARPS-N spectra (with resulting S/N = 96 at 550~nm)
was analysed using the same procedures described in detail by \citet{Sozzettietal2004, Sozzettietal2006}
and \citet{Dumusqueetal2014}, and references therein. A first set of relatively weak Fe I and 
Fe II lines was selected from the \citet{Sousaetal2010} list, and equivalent 
widths (EWs) were measured using the TAME software \citep{KangLee2012}. A second set of iron lines 
was chosen from the list of \citet{Biazzoetal2012}, and EWs were measured manually. 
Effective temperature $T_\mathrm{eff}$, surface gravity $\log g$, microturbulence velocity $\xi_t$, 
and iron abundance [Fe/H] were then derived under the assumption of local thermodynamic equilibrium (LTE), 
using the 2013 version of the spectral synthesis code MOOG \citep{Sneden1973}, a grid of Kurucz ATLAS 
plane-parallel model stellar atmospheres \citep{Kurucz93}, and by imposing excitation and ionisation 
equilibrium. Within the error bars, the two methods provided consistent results. 
The final adopted values, obtained as the weighted mean of the two independent determinations, 
are summarized in Table~\ref{starplanet_param_table}. They agree, to within $1~\sigma$, with
the atmospheric parameters found independently with SPC \citep{Buchhaveetal2014}.
Both the low $V \sin{i_{*}}=2.6 \pm 0.5~\kms$ and the average activity index $\log{R^{'}_{HK}}=-5.17 \pm 0.05$ 
further support the low magnetic activity level of the host star inferred from the \emph{Kepler} light curve.
  
\section{Data analysis and system parameters} 
\label{sys_param} 
To determine the system parameters, a Bayesian combined analysis of 
HARPS-N and \emph{Kepler} data was performed using 
a Differential Evolution Markov Chain Monte Carlo (DE-MCMC) method 
\citep{TerBraak2006, Eastmanetal2013}. 
SC \emph{Kepler} data were used 
to model the transits of Kepler-101b, because they yield 
a more accurate solution than LC measurements by avoiding 
the distortion of the transit shape caused by the LC sampling \citep{Kipping10}.
Specifically, eighty-two transits of Kepler-101b were observed in SC mode, which
yields a S/N of $\sim 190$ for the phase-folded transit.
On the other hand, LC measurements were used to 
perform the Kepler-101c transit modeling 
because SC data alone do not provide a high enough S/N. Indeed,
fifty transits of Kepler-101c were observed with SC sampling, while 
two-hundred and twenty-nine in LC mode.
To perform the transit fitting, transits of Kepler-101b and c were individually normalized 
by fitting a linear function of time to the light curve intervals of twice 
the transit duration before their ingress and after their egress. 

%Correlated noise was estimated following \citet{Pontetal2006} and \citet{Bonomoetal2012}
%%and added in quadrature to the formal error bars although
%but its contribution was found to be practically negligible.

Since the RV signal of Kepler-101c is not detected in our HARPS-N data, 
we first performed a combined 
analysis of \emph{Kepler} photometry and HARPS-N radial velocities of Kepler-101b
by fitting simultaneously a transit model \citep{Gimenez06} and a Keplerian orbit.
The free parameters of our global model are 
the transit epoch $T_{\rm 0}$, the orbital period $P$, 
two systemic radial velocities for HARPS-N data obtained with both the original ($V_{\rm r,o}$) and the 
replaced chip ($V_{\rm r,r}$), 
the radial-velocity semi-amplitude $K$, 
$\sqrt{e}~{\cos{\omega}}$ and  
$\sqrt{e}~{\sin{\omega}}$, where $e$  is the eccentricity and
$\omega$ the argument of periastron, 
the transit duration $T_{\rm 14}$,
the ratio of the planet to stellar radii $R_{\rm p}/R_{*}$, 
the inclination $i$ between the orbital plane and the plane of the sky,
and the two limb-darkening coefficients (LDC)
$q_{1}=(u_{a}+u_{b})^2$ and $q_{2}=0.5 u_{a} / (u_{a}+u_{b})$ \citep{Kipping2013}, 
where $u_{\rm a}$ and $u_{\rm b}$ are the coefficients of the quadratic limb-darkening law (e.g., \citealt{Claret2000}).
A DE-MCMC analysis with a number of chains equal to twice the number of free parameters
was then carried out. After removing the ``burn-in'' steps, as suggested by \citet{Knutsonetal2009}, and 
achieving convergence and well-mixing of the chains according to \citet{Ford2006}, 
the medians of the posterior distributions and their $34.13\%$ intervals
were evaluated and were taken as the final parameters and 
associated $1~\sigma$ uncertainties, respectively.
Mass, radius, and age of the host star were determined by 
comparing the Yonsei-Yale evolutionary tracks \citep{Demarqueetal2004}
with the stellar effective temperature, metallicity, and 
density as derived from $a/R_\star$ and Kepler's third law 
(see, e.g., \citealt{Sozzettietal2007, Torresetal2012}).
For this purpose, we considered normal distributions for the $T_{\rm eff}$ and [Fe/H] with
standard deviations equal to the uncertainties derived from our spectral analysis (Sect.~\ref{star_atmos_param}).
We used the same chi-square minimization as described in \citet{Santerneetal2011}. 
%The stellar density, as derived from $a/R_\star$ and Kepler's third law, 
%was used as a proxy for stellar luminosity to determine
%the mass, radius, and age of the host star \citep{Sozzettietal2007, Torresetal2012},
%given its effective temperature and metallicity (Sect.~\ref{star_atmos_param}). 
Orbital, stellar, and Kepler-101b parameters are reported in Table~\ref{starplanet_param_table}.
The SC photometric measurements and HARPS-N data phase-folded with the 
ephemeris of Kepler-101b are shown in Fig.~\ref{fig_Kepler101b}
along with the best solution.

The parent star Kepler-101 is a slightly evolved and metal-rich star, with a mass of 
$1.17_{-0.05}^{+0.07}~\Msun$, a radius of $1.56 \pm 0.20~\Rsun$, and an 
age of $5.9 \pm 1.2$~Gyr. According to orbital, transit, 
and the above-mentioned stellar parameters, Kepler-101b has mass, radius, and density of 
$\mp=51.1_{-4.7}^{+5.1}~\Me$, $\rp=5.77_{-0.79}^{+0.85}~\Re$, and 
$\rho_{\rm p}=1.45_{-0.48}^{+0.83}~\rm g\;cm^{-3}$. 
These values of mass and radius are in-between those of Neptune and Saturn,
making Kepler-101b deserve the name of ``super-Neptune". 
Given its vicinity to the host star ($a=0.047$~au), the equilibrium temperature is high, i.e. $T_{\rm eq}\sim1515$~K.
The inferred eccentricity of Kepler-101b is consistent with zero, although - given the current precision - a 
small eccentricity ($< 0.17$ at $1~\sigma$) can not be excluded.

\begin{figure}[h]
\centering
%\begin{minipage}{15cm}
%\vspace{0.5cm}
\includegraphics[width=6.5cm, angle=90.]{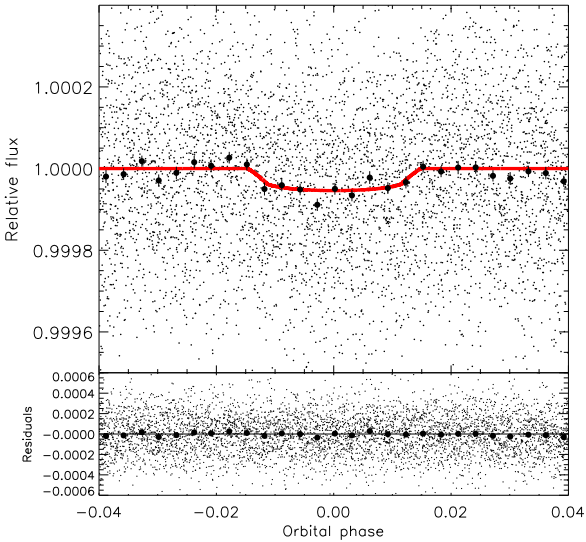}
%\hspace{1.0cm}
%\includegraphics[width=7cm]{plot_rv_KOI46_tdb.eps}
\vspace{0.5cm}
\caption{Planetary transit of the Earth-sized planet Kepler-101c with the transit model (red solid line). 
Small circles show the phase-folded long-cadence \emph{Kepler} data. Larger circles are the same 
data binned in 0.003 phase intervals for display purpose.
}
\label{fig_Kepler101c}
%\end{minipage}
\end{figure}

The physical parameters of Kepler-101c were determined after those of Kepler-101b. The modeling of its transit was carried out by i) using all the
available LC data, ii) considering a circular orbit\footnote{a circular orbit was adopted 
for Kepler-101c in the absence of any RV constraint on orbital eccentricity. The latter, in any case,
must be lower than 0.2 so as to avoid orbit crossing and system instability. See Sect.~\ref{discussion_conclusion}},
iii) oversampling the transit model by a factor of 10 (e.g., \citealt{Southworth2012}), 
and iv) fixing the LDC to those which were previously derived 
because the low transit S/N prevents us from fitting them.  
The transit of Kepler-101c along with the best fit is shown in Fig.~\ref{fig_Kepler101c}.
A Bayesian DE-MCMC combined analysis of \emph{Kepler} photometry 
and the residuals of HARPS-N data, after subtracting the Kepler-101b signal, 
indicates that this planet has a radius of $1.25_{-0.17}^{+0.19}~\Re$ and 
a mass $< 3.78~\Me$. Indeed, as already mentioned, only an upper limit 
on the RV semi-amplitude of Kepler-101c can be given, i.e. $K < 1.17~\ms$. 
Consequently, its bulk density is highly uncertain: $\rho_{\rm p}< 10.5~\rm g\;cm^{-3}$.
Finally, we point out that both the upper limit on the RV semi-amplitude of Kepler-101c 
and the orbital parameters of Kepler-101b were found to be fully consistent 
when modeling the RV data with two Keplerian orbits and imposing gaussian priors on the orbital 
periods and transit epochs of Kepler-101b and c from \emph{Kepler} photometry.

\begin{table}
%\vspace{-0.4cm}
\centering
\caption{Kepler-101 system parameters. Errors and upper limits refer to $1~\sigma$ uncertainties.}            
%\vspace{1cm}
\begin{minipage}[t]{9.0cm} 
\setlength{\tabcolsep}{1.0mm}
\renewcommand{\footnoterule}{}                          
\begin{tabular}{l l}        
%\hline  
%\\
\hline\hline
\emph{Stellar IDs, coordinates, and magnitudes} &  \\
\hline
\emph{Kepler} ID & 10905239 \\
\emph{Kepler} Object of Interest & KOI-46 \\
USNO-A2 ID & 1350-09997781 \\
2MASS ID & 18530131+4821188 \\
RA (J2000)   & 18:53:01.32  \\
Dec (J2000) &  48:21:18.84  \\
\emph{Kepler} magnitude & 13.77 \\
Howell Everett Survey Johnson-B & 14.52 \\
Howell Everett Survey Johnson-V & 13.80 \\
2MASS J & $12.40 \pm 0.02$ \\
%2MASS H & $12.11 \pm 0.02$ \\
2MASS K & $12.01 \pm 0.02$ \\
\hline\hline
\emph{Stellar parameters} &  \\
\hline
%Effective temperature $T_{\rm{eff}}$[K] & \multicolumn{2}{c}{5667 $\pm$ 50} \\
Effective temperature $T_{\rm{eff}}$[K]& 5667 $\pm$ 50 \\
Metallicity $[\rm{Fe/H}]$ [dex] & 0.33  $\pm$ 0.07 \\
Microturbulence velocity $\xi_{t}$ [\kms] & $1.00 \pm 0.05$ \\
Rotational velocity $V \sin{i_{*}}$ [\kms] & 2.6 $\pm$ 0.5 \\
Density $\rho_{*}$ [$ \rm g\;cm^{-3}$] & $0.437_{-0.123}^{+0.204}$ \\
Mass [\Msun] & $1.17_{-0.05}^{+0.07}$ \\
Radius [\Rsun] & $1.56 \pm 0.20$  \\
Derived surface gravity log\,$g$ [cgs] & $4.12_{-0.09}^{+0.11}$ \\
Age $t$ [Gyr] & $5.9 \pm 1.2$ \\
Spectral type & G3IV  \\
%Limb-darkening coefficient $q_{1}$  & $0.53_{-0.11}^{+0.14}$ \\
%Limb-darkening coefficient $q_{2}$  & $0.19_{-0.09}^{+0.12}$  \\
Linear limb-darkening coefficient $u_{a}$  & $0.28 \pm 0.13$ \\
Quadratic limb-darkening coefficient $u_{b}$  & $0.46 \pm 0.20$ \\
\hline\hline
\emph{Kepler-101\,b}  &   \\
\hline
\emph{Transit and orbital parameters} &    \\
Orbital period $P$ [days] & 3.4876812 $\pm$ 0.0000070\\
Transit epoch $T_{ \rm 0} [\rm BJD_{TDB}-2454900$] & 288.77995 $\pm$ 0.00041 \\
Transit duration $T_{\rm 14}$ [h] & $3.875_{-0.020}^{+0.023}$ \\
Radius ratio $R_{\rm p}/R_{*}$ & $0.03401_{-0.00082}^{+0.00061}$  \\
Inclination $i$ [deg] & $85.82_{-1.53}^{+1.73}$ \\
$\sqrt{e}~\cos{\omega}$ & $ -0.13_{-0.11}^{+0.13}$ \\
$\sqrt{e}~\sin{\omega}$ & $ -0.17_{-0.21}^{+0.26} $ \\
Orbital eccentricity $e$  & $0.086_{-0.059}^{+0.080} $  \\
Argument of periastron $\omega$ [deg] & $231_{-90}^{+32}$  \\
Radial velocity semi-amplitude $K$ [\ms] & $19.4 \pm 1.8 $ \\
Systemic velocity $V_{\rm r, r}$ [\kms] & $-77.7110 \pm 0.0015$ \\
Systemic velocity $V_{\rm r, o}$ [\kms] & $-77.7440 \pm 0.0025$ \\
$a/R_{*}$ & $6.55_{-0.69}^{+0.88}$ \\
%Stellar density $\rho_{*}$ [$ \rm g\;cm^{-3}$] & $0.368_{-0.069}^{+0.112}$ & $0.435_{-0.118}^{+0.210}$ \\
Impact parameter $b$ & $0.52_{-0.18}^{+0.09}$ \\
%& \\
\hline
\emph{Planetary parameters} &  \\
Mass $M_{\rm p} ~[\rm M_\oplus]$  & $51.1_{-4.7}^{+5.1}$ \\
Radius $R_{\rm p} ~[ \rm R_\oplus]$  & $5.77_{-0.79}^{+0.85}$ \\
Density $\rho_{\rm p}$ [$\rm g\;cm^{-3}$] & $1.45_{-0.48}^{+0.83} $ \\
Surface gravity log\,$g_{\rm p }$ [cgs] & $3.17_{-0.11}^{+0.13}$ \\
Orbital semi-major axis $a$ [au] & $0.0474_{-0.0008}^{+0.0010}$  \\
Equilibrium temperature $T_{\rm eq}$ [K] ~$^a$ & $1513_{-145}^{+103}$\\
\hline\hline       
\emph{Kepler-101\,c}  &  \\
\hline
\emph{Transit and orbital parameters} &  \\
Orbital period $P$ [days] & 6.029760 $\pm$ 0.000075   \\
Transit epoch $T_{ \rm 0} [\rm BJD_{TDB}-2454900$] &  $65.4860 \pm 0.0088$    \\
Transit duration $T_{\rm 14}$ [h] & $3.87 \pm 0.24$   \\
Radius ratio $R_{\rm p}/R_{*}$ & $0.00732_{-0.00054}^{+0.00063}$    \\
Inclination $i$ [deg] & $84.6_{-3.1}^{+3.4}$  \\
Orbital eccentricity $e$  & 0 (fixed)  \\
Argument of periastron $\omega$ [deg] & 90 (fixed)  \\
Radial velocity semi-amplitude $K$ [\ms] & $ < 1.17 $   \\
$a/R_{*}$ & $8.0_{-2.1}^{+3.0}$ \\
Impact parameter $b$ & $0.75_{-0.36}^{+0.13}$  \\
\hline
\emph{Planetary parameters} &   \\
Mass $M_{\rm p} ~[\rm M_\oplus]$  &  $ < 3.78 $  \\
Radius $R_{\rm p} ~[ \rm R_\oplus]$  &  $1.25_{-0.17}^{+0.19}$  \\
Density $\rho_{\rm p}$ [$\rm g\;cm^{-3}$] &  $ < 10.5 $  \\
%Planet surface gravity log\,$g_{\rm p }$ [cgs] &  $3.12 \pm 0.09$ &  -  \\
Orbital semi-major axis $a$ [au] & $0.0684 \pm 0.0014$  \\
Equilibrium temperature $T_{\rm eq}$ [K] ~$^a$ & $1413_{-210}^{+238}$ \\
\hline\hline
\vspace{-0.5cm}
\footnotetext[1]{\scriptsize Black-body equilibrium temperature assuming a null Bond albedo and uniform 
heat redistribution to the night side.} \\
\end{tabular}
\end{minipage}
\label{starplanet_param_table}  
\end{table}

%~\\

\begin{table}
%\vspace{-0.4cm}
\centering
\caption{HARPS-N radial velocities and bisector spans of Kepler-101.
The last column indicates the HARPS-N CCD: O and R stand for the original and 
the replaced CCD, respectively (see text).}            
%\vspace{1cm}
%\begin{minipage}[t]{9.0cm} 
\setlength{\tabcolsep}{1.0mm}
\renewcommand{\footnoterule}{}                          
\begin{tabular}{c c c c c}        
\hline \hline
$ \rm BJD_{UTC}$ & RV & $\pm 1~\sigma$ & Bis. span & CCD \\
-2450000   & ($\ms$) & ($\ms$) & ($\ms$) &  \\
\hline
6102.618403  & -77739.82  &   8.38  &  -28.13 & O	\\
6102.639630  & -77757.71  & 10.60  &  -35.69 & O	\\
6114.521305  & -77764.50  & 11.65  &  -50.76 & O	\\
6114.542544  & -77743.52  & 12.18  &  -52.91 & O	\\
6115.514333  & -77727.63  &   6.28  &  -8.35 & O	\\
6115.535548  & -77729.12  &   5.84  &  -35.93 & O	\\
6116.514941  & -77739.54  &   7.98  &  -25.19 & O	\\
6116.536133  & -77739.84  &   7.34  &  -13.34 & O 	\\
6117.475560  & -77764.48  &   6.31  &  -10.70 & O	\\
6117.496811  & -77762.48  &   6.53  &  -20.69 & O	\\
6436.576753  & -77690.77  &   5.48  &  -20.76 & R	\\
6437.676342  & -77732.91  &   7.26  &    -9.82 & R	\\
6462.515513  & -77733.73  &   8.71  &  -35.87 & R	\\
6463.465239  & -77701.82  &   9.73  &  -24.03 & R	\\
6482.571731  & -77707.87  & 11.48  &   16.11 & R		\\
6482.593676  & -77713.09  & 11.15  &  -25.66 & R	\\
6495.416309  & -77696.42  &   5.96  &     1.10 & R		\\
6497.548192  & -77737.89  &   9.71  &     5.84 & R		\\	
6497.564893  & -77735.63  &   8.70  &  -15.85 & R	\\
6498.638143  & -77695.98  &   9.68  &  -38.55 & R	\\
6498.652622  & -77697.74  & 12.27  &  -54.50 & R	\\
6499.530536  & -77686.68  &   9.26  &   16.28 & R 	\\
6499.545593  & -77688.48  &   8.77  &   29.85 & R		\\
6500.412431  & -77717.32  &   7.05  &  -25.93 & R	\\
6500.428599  & -77729.78  &   7.32  &  -35.69 & R	\\
6501.556731  & -77732.92  & 10.31  &  -63.39 & R	\\
6510.450025  & -77698.01  &   8.45  &  -10.25 & R	\\
6511.445178  & -77730.92  &   7.26  &    -1.55 & R	\\
6511.466532  & -77725.08  &   7.87  &  -65.89 & R	\\
6512.447436  & -77701.43  &   6.46  &  -21.48 & R	\\
6512.468362  & -77702.63  &   6.08  &  -12.55 & R	\\
6513.439347  & -77702.26  &   6.73  &     8.23 & R		\\
6513.460087  & -77704.35  &   5.80  &  -52.45 & R	\\
6514.389197  & -77716.24  &   8.60  &     1.79  & R	\\
6514.410123  & -77715.08  &   8.74  &  -27.32 & R	\\
6515.380077  & -77727.62  & 10.05  & -13.66  & R	\\
6515.401315  & -77718.09  & 11.59  &   16.72  & R	\\
6528.437262  & -77706.54  & 10.10  &  -30.15 & R	\\
6530.406773  & -77690.25  &   6.29  &    -2.74 & R	\\
6532.371144  & -77732.99  &   7.62  &  -17.58 & R	\\
\hline \hline
\label{table_rv}
\end{tabular}
\end{table}

\section{Discussion and conclusions}
\label{discussion_conclusion}
Thanks to forty precise spectroscopic observations obtained with
HARPS-N, and a Bayesian combined analysis of these measurements and 
\emph{Kepler} photometry, we were able to characterize 
the Kepler-101 planetary system. The system consists of a hot super-Neptune, 
Kepler-101b at a distance of 0.047~au from the host star, and an outer
Earth-sized planet, Kepler-101c with semi-major axis of 0.068~au
and mass $< 3.8~\Me$. 

\begin{figure}
\centering
%\begin{minipage}{15cm}
%\vspace{0.5cm}
\includegraphics[width=9cm]{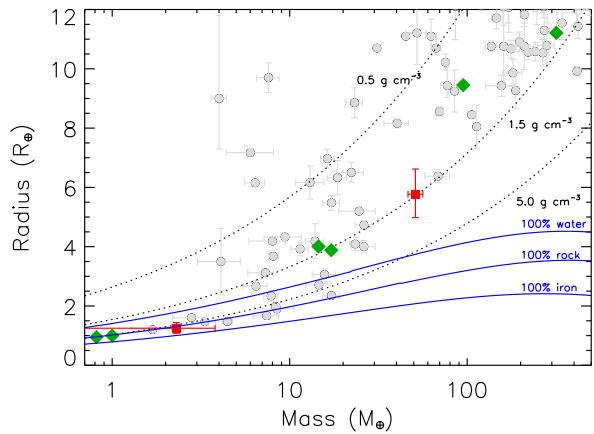}
%\includegraphics{massradius_kepler-101.ps}
%\hspace{1.0cm}
%\includegraphics[width=7cm]{plot_rv_KOI46_tdb.eps}
%\vspace{0.5cm}
\caption{Mass-radius diagram of the known transiting planets with radius $\rp \leq 12~\Re$, 
mass $\mp < 500~\Me$, and precision on the mass better than 30~\%. 
The three dotted lines correspond to different isodensity
curves. From top to bottom, the blue solid lines indicate mass and radius for planets consisting in pure water,
100\% silicates, and 100\% iron \citep{Seageretal2007}. The positions of Kepler-101b and c are plotted with red squares.
}
\label{fig_massradius}
%\end{minipage}
\end{figure}

Figure~\ref{fig_massradius} shows the positions of Kepler-101b and c 
in the radius-mass diagram of known exoplanets 
with radius $\rp \leq 12~\Re$, mass $\mp < 500~\Me$, 
and precision on the mass better than 30\%. Green diamonds
indicate the Solar System giant planets Jupiter, Saturn, Neptune, 
Uranus, and the terrestrial planets Earth and Venus (from right to left). 
The three dotted lines indicate isodensity
curves of 0.5, 1.5, and 5 $\rm g\,cm^{-3}$ (from top to bottom), and 
the blue solid lines show the mass and radius of planets consisting of 
pure water, 100\% rocks, and 100\% iron \citep{Seageretal2007}. 
Kepler-101b joins the rare known transiting planets in the transition region between 
Saturn-like and Neptune-like planets. A lower occurrence of giant planets in the mass interval
$30 \lesssim M_{\rm p} \lesssim 70~\Me$ is expected from 
certain models of planet formation through core accretion followed by planetary migration
and disc dissipation (e.g., \citealt{Mordasinietal2009}; see their Fig.~3).
This is likely related to the fact that when a protoplanet reaches the critical mass
to undergo runaway accretion \citep{Pollacketal1996}, its mass quickly increases up to 1-3~$\Mjup$.
Therefore, disc dissipation might have occurred at the time of gas supply \citep{Mordasinietal2009}.
Indeed, the X-ray and EUV energy flux from the parent star 
can account for a mass loss of $<13~\Me$ during its lifetime of $\sim 6$~Gyr,
according to \citet{Sanz-Forcadaetal2011}.

%can account for a change in the planetary mass $<1~\Me$ during its lifetime of $\sim 6$~Gyr,
%according to Eq.~15 in \citet{LecavelierdesEtangs2007}.

In terms of mass and radius, Kepler-101b is, to our knowledge, the first fully-characterized super-Neptune. 
Indeed, seven known transiting planets with accurately measured masses 
have radii comparable with that of Kepler-101b at $1~\sigma$, 
namely CoRoT-8b, HAT-P-26b, Kepler-18c, Kepler-25b, Kepler-56b, Kepler-87c, and Kepler-89e.  
However, CoRoT-8b is a dense sub-Saturn 
with a higher mass than Kepler-101b, i.e. $M_{\rm p}=69.9 \pm 9.5~\Me$ \citep{Bordeetal2010}, 
and the remaining planets are low-density Neptunes with masses below $\sim 25~\Me$.
All the three planets with mass comparable to that of Kepler-101b mass, i.e. Kepler-35b, 
Kepler-89d, and Kepler-9b, have larger radii, i.e. $\rp > 8~\Re$.

According to models of the internal structure of irradiated \citep{Fortneyetal2007, LopezFortney2013} 
and non-irradiated \citep{Mordasinietal2012} planets, 
Kepler-101b's interior should contain a significant amount of heavy elements; more than 60\% of its total mass. %(see Fig.~9 in \citealt{Mordasinietal2012}).
This might further support the observed correlation between the heavy element
content of giant planets and the metallicity of their parent stars \citep{Guillot2008}. 
Detailed modeling of the internal structure of the Earth-sized planet Kepler-101c is not 
possible because of the weak constraint on its mass, i.e. $M_{\rm p}< 3.8~\Me$ ($< 8.7~\Me$)
at 1~$\sigma$ (2~$\sigma$). We are only able to exclude a composition of pure iron with 
$68.3\%$ probability, according to models for solid planets 
by \citet{Seageretal2007} and \citet{ZengSasselov2013}, and 
any H/He envelope from the planetary radius constraint \citep{Rogersetal2011}.

We carried out a small number of N-body runs using a Hermite integrator \citep{Makino1991}, to investigate the 
stability of the Kepler-101 planetary system. The mass of the star and inner planet were set 
according to the values in Table~\ref{starplanet_param_table},
as were the semi-major axis and eccentricity of the inner planet.  The semi-major axis of the outer planet
was also set according to the value in Table~\ref{starplanet_param_table}, 
but we varied its mass and the initial eccentricity of its orbit.
We ran each simulation for at least $10^8$ orbits of the inner planet. Our simulations suggest that the system
is stable for outer planet masses between 1 and 4 Earth masses and for outer planet eccentricities $e \le 0.2$. 
Orbit crossing would occur if the outer eccentricity exceeded $e = 0.25$ and 
%Although we didn't determine precisely at what eccentricity the system would become unstable, 
none of our simulations with an outer planet eccentricity $e \ge 0.225$ was stable. 
Hence, our results indicate that 
the system is stable for masses $\lesssim 4~\Me$ of the outer planet, in agreement with the 
$1~\sigma$ upper limit from RV measurements, and for all eccentricities $e \lesssim 0.2$.

The fact that both planets in the Kepler-101 system transit their parent star 
suggests that these planets evolved through disc-planet interactions \citep{kley12},
rather than through dynamical interactions \citep{rasio96}.  One suggested
mechanism for forming close-in multiple planets is that differential migration
forces the planets into a stable resonance, after which the planets migrate
inwards together \citep{lee02}. That a large fraction of the known multiple
planet systems are in mean motion resonance (MMR) \citep{crida08} would seem to be
consistent with this picture. The Kepler-101 planets, however, are not 
in a mean motion resonance, and so this seems an unlikely formation
scenario. On the other hand, the inner planet is close enough to the host star 
that it is likely that its orbit is influenced by a tidal interaction. 
Therefore, it is possible that the two planets did migrate inwards in a 3:2 MMR 
but that the inner planet has since migrated further inwards 
due to tidal effects \citep{DelisleLaskar2014}. %, 
%the outer planet having an orbit sufficiently far from the host star that 
%tidal effects will not be significant. 
Additionally, neither planet is massive enough to undergo
gap opening Type II migration \citep{lin86}, and so they would be expected
to migrate in the faster Type I regime \citep{ward97}. That the inner planet
is more massive than the outer, would also suggest that differential migration 
should cause these planets to separate, rather than migrating into a resonant 
configuration. Additionally, the density of Kepler-101b suggests that it
likely formed beyond the snowline.  Although we don't have an accurate density
for Kepler-101c, the data may suggest that it has a composition consistent
with formation inside the snowline.  If so, this may be an example of a
system in which an inner planet has survived the passage of a more massive
outer planet, been scattered onto a wider orbit, and then migrated inwards 
to where it is today \citep{fogg05,cresswell08}. 
Indeed, the Kepler-101 system does not follow the trend observed for 
$\sim 70\%$ of \emph{Kepler} planet pairs with at least 
one Neptune-size or larger planet. In such systems, 
the larger planet, typically, has the longer period \citep{Ciardietal2013}.

In conclusion, both the architecture of the Kepler-101 planetary system 
and the first full characterization of a super-Neptune % Kepler-101b
are certainly of interest for a better understanding of 
planet formation and evolution, and for the study of the internal structures of 
giant planets in the transition region between Saturn-like and Neptune-like planets.

\begin{acknowledgements}
The HARPS-N project was funded by the Prodex Program of the Swiss Space Office (SSO), the Harvard- University Origin of Life Initiative (HUOLI), the Scottish Universities Physics Alliance (SUPA), the University of Geneva, the Smithsonian Astrophysical Observatory (SAO), and the Italian National Astrophysical Institute (INAF), University of St. Andrews, QueenÕs University Belfast and University of Edinburgh. The research leading to these results has received funding from the European Union Seventh Framework Programme (FP7/2007-2013) under Grant Agreement n. 313014 (ETAEARTH). 
This research has made use of the results produced by the PI2S2 Project managed by the Consorzio COMETA, a co-funded project by the Italian Ministero dell'Istruzione, Universit\`a e Ricerca (MIUR) within the Piano Operativo Nazionale Ricerca Scientifica, Sviluppo Tecnologico, Alta Formazione (PON 2000Ð2006).
X. Dumusque would like to thank the Swiss National Science Foundation (SNSF) for its support through an Early Postdoc Mobility fellowship. P. Figueira acknowledges support by  Funda\c{c}\~ao para a Ci\^encia e a Tecnologia (FCT) through the Investigador FCT contract of reference IF/01037/2013 and POPH/FSE (EC) by FEDER funding through the program ``Programa Operacional de Factores de Competitividade - COMPETE''. 
R.D. Haywood acknowledges support from an STFC postgraduate research studentship.
This publication was made possible through the support of a grant from the John Templeton Foundation. The opinions expressed in this publication are those of the authors and do not necessarily reflect the views of the John Templeton Foundation.
%We would like to thank A. McWilliam, I. Ivans and C. Sneden for providing us their software that interpolates between atmospheric models. 
\end{acknowledgements}


\begin{thebibliography}{}
\bibliographystyle{aa}

{\small

%\bibitem[\protect\citeauthoryear{Anderson et al.} {2011}]{Andersonetal2011}
%Anderson, D. R., Collier Cameron, A., Hellier, C. et al. 2011, ApJ, 726, L19

\bibitem[\protect\citeauthoryear{Baranne} {1996}]{baranne96} 
Baranne, A., Queloz, D., Mayor, M., et al. 1994, \aaps, 119, 373

%\bibitem[\protect\citeauthoryear{Batalha et al.}{2013}]{Batalhaetal2013}
%Batalha, N., Rowe, J. F., Bryson, S. T., et al. 2013, \apjs, 204, 24

%\bibitem[\protect\citeauthoryear{Bonomo et al.} {2010}]{Bonomoetal2010} 
%Bonomo, A. S., Santerne, A., Alonso, R., et al. 2010 \aap, 520, A65

%\bibitem[\protect\citeauthoryear{Bonomo et al.} {2012b}]{Bonomoetal2012}
%Bonomo, A., S.,  Chabaud, P.-Y., Deleuil, M. et al. 2012b, \aap, 547, A110


\bibitem[\protect\citeauthoryear{Biazzo et al.} {2012}]{Biazzoetal2012}	
Biazzo, K., D'Orazi, V., Desidera, S., et al. 2012, \mnras, 427, 2905

\bibitem[\protect\citeauthoryear{Bord\'e et al.} {2010}]{Bordeetal2010}
Bord\'e, P., Bouchy, F., Deleuil, M., et al. 2010, \aap, 520, A66

\bibitem[\protect\citeauthoryear{Buchhave et al.} {2014}]{Buchhaveetal2014}
Buchhave, L., Bizarro, M., Latham, D. W., et al. 2014, Nature, 509, 593

\bibitem[\protect\citeauthoryear{Ciardi et al.} {2013}]{Ciardietal2013}
Ciardi, D. R., Fabrycky, D. C., Ford, E. B. et al. 2013, \apj, 763, 41

\bibitem[\protect\citeauthoryear{Claret} {2000}]{Claret2000}
Claret, A. 2000, \aap, 363, 1081

%\bibitem[\protect\citeauthoryear{Cochran et al.} {2011}]{Cochranetal2011}
%Cochran, W. D., Fabrycky, D. C., Torres, G. et al. 2011, \apjs, 197, 7

\bibitem[\protect\citeauthoryear{Cosentino et al.} {2012}]{Cosentinoetal2012}
Cosentino, R., Lovis, C., Pepe, F., et al. 2012, in Society of Photo-
Optical Instrumentation Engineers (SPIE) Conference Series, Vol. 8446, 
Society of Photo-Optical Instrumentation Engineers (SPIE)
Conference Series

%\bibitem{cresswell08}
\bibitem[\protect\citeauthoryear{Cresswell \& Nelson}{2008}]{cresswell08}
Cresswell, P., \& Nelson, R.P. 2008, A\&A, 482, 677

%\bibitem{crida08}
\bibitem[\protect\citeauthoryear{Crida et al.}{2008}]{crida08}
Crida, A., Sandor, Z., \& Kley W. 2008, A\&A, 483, 325

\bibitem[\protect\citeauthoryear{Delisle \& Laskar} {2014}]{DelisleLaskar2014}
Delisle, J.-B., \& Laskar, J. 2014, \aap, accepted, arXiv:1406.0694

\bibitem[\protect\citeauthoryear{Demarque et al.} {2004}]{Demarqueetal2004}
Demarque, P., Woo, J.-H., Kim, Y.-C., \& Yi, S. K. 2004, \apjs, 155, 667

\bibitem[\protect\citeauthoryear{Dumusque et al.} {2014}]{Dumusqueetal2014}
Dumusque, X., Bonomo, A. S., Haywood, R. D., et al. 2014, \apj, accepted, arXiv:1405.7881

%\bibitem[\protect\citeauthoryear{Eastman et al.} {2010}]{Eastmanetal2010}
%Eastman, J., Siverd, R., \& Gaudi, B. S. 2010, \pasp, 122, 935

\bibitem[\protect\citeauthoryear{Eastman et al.} {2013}]{Eastmanetal2013}
Eastman, J., Gaudi, B. S., \& Agol, E. 2013, \pasp, 125, 923

%\bibitem[\protect\citeauthoryear{Ford} {2005}]{Ford2005}
%Ford, E. B. 2005, \aj, 129, 1706

%\bibitem{fogg05}
\bibitem[\protect\citeauthoryear{Fogg \& Nelson}{2005}]{fogg05}
Fogg, M. J., \& Nelson, R.P. 2005, A\&A, 441, 791

\bibitem[\protect\citeauthoryear{Fogg \& Nelson}{2007}]{FoggNelson2007}
Fogg, M. J., \& Nelson, R. P. 2007, \aap, 461, 1195

\bibitem[\protect\citeauthoryear{Ford} {2006}]{Ford2006}
Ford, E. B. 2006, \apj, 642, 505

\bibitem[\protect\citeauthoryear{Fortney et al.} {2007}]{Fortneyetal2007}
Fortney, J. J., Marley, M. S., \& Barnes, J. W. 2007, \apj, 659, 1661

%\bibitem[\protect\citeauthoryear{Gelman et al.} {2004}]{Gelmanetal2004}
%Gelman A., Carlin J. B., Stern H. S. and Rubin D. B. 2004. Bayesian
%data analysis, 2nd edition. London, Chapman \& Hall.

\bibitem[\protect\citeauthoryear{Gim\'enez} {2006}]{Gimenez06}
Gim\'enez, A. 2006, \aap, 450, 1231

\bibitem[\protect\citeauthoryear{Guillot} {2008}]{Guillot2008}
Guillot, T. 2008, Physical Scripta, 130, id. 014023

%\bibitem[\protect\citeauthoryear{Hartman et al.} {2011}]{Hartmanetal2011} 
%Hartman, J. D., Bakos, G. \'A., Kipping, D. M., et al. 2011, \apj, 728, 138

%\bibitem[\protect\citeauthoryear{Huber et al.} {2013}]{Huberetal2013} 
%Huber, D., Carter, J. A., Barbieri, M., et al. 2013, Science, 342, 331

\bibitem[\protect\citeauthoryear{Ida \& Lin}{2008}]{IdaLin2008}
Ida, S., \& Lin, D. N. C. 2008, \apj, 673, 487

\bibitem[\protect\citeauthoryear{Jenkins et al.} {2010}]{Jenkinsetal2010} 
Jenkins, J. M., Caldwell, D. A., Chandrasekaran, H. et al. 2010, \apj, 713, L87

\bibitem[Kang \& Lee, 2012]{KangLee2012}
Kang, W., \& Lee, S.-G. 2012, \mnras, 425, 3162

\bibitem[\protect\citeauthoryear{Kipping} {2010}]{Kipping10} 
Kipping, D. 2010, \mnras, 408, 1758

\bibitem[\protect\citeauthoryear{Kipping} {2013}]{Kipping2013} 
Kipping, D. 2013, \mnras, 435, 2152

%\bibitem{kley12}
\bibitem[\protect\citeauthoryear{Kley \& Nelson}{2012}]{kley12}
Kley, W., \& Nelson, R.P. 2012, ARA\&A, 50, 211

\bibitem[\protect\citeauthoryear{Knutson et al.} {2009}]{Knutsonetal2009} 
Knutson, H. A., Charbonneau, D., Cowan, N. B., et al. 2009, \apj, 703, 769

\bibitem[\protect\citeauthoryear{Kurucz} {1993}]{Kurucz93} 
Kurucz, R. I. 1993, ATLAS9 Stellar Atmosphere Programs and 2 km/s grid. 
Kurucz CD-ROM No. 13. Cambridge, Mass.: Smithsonian Astrophysical Observatory, 1993.

\bibitem[\protect\citeauthoryear{Latham et al.} {2011}]{Lathametal2011}
Latham, D. W., Rowe, J. F., Samuel, N. Q., et al. 2011, \apj, 732, L24

%\bibitem[\protect\citeauthoryear{Lecavelier des Etangs}{2007}]{LecavelierdesEtangs2007}
%Lecavelier des Etangs 2007, \aap, 461, 1185

%\bibitem{lee02}
\bibitem[\protect\citeauthoryear{Lee \& Peale}{2002}]{lee02}
Lee, M.H., \& Peale, S. J. 2002, ApJ, 567, 596

%\bibitem{lin86}
\bibitem[\protect\citeauthoryear{Lin \& Papaloizou}{1986}]{lin86}
Lin, D.N.C., \& Papaloizou, J. 1986, ApJ, 307, 395

\bibitem[\protect\citeauthoryear{Lopez \& Fortney}{2013}]{LopezFortney2013}
Lopez, E. D., \& Fortney J. J. 2013, submitted, arXiv:1311.0329 

\bibitem[\protect\citeauthoryear{Makino}{1991}]{Makino1991}
Makino, J. 1991, \apj, 369, 200

\bibitem[\protect\citeauthoryear{Mayor et al.}{2003}]{Mayoretal2003}	
Mayor, M., Pepe, F., Queloz, D., et al. 2003, The Messenger (ISSN0722-6691), No.114, p. 20-24

\bibitem[\protect\citeauthoryear{Mordasini et al.}{2009}]{Mordasinietal2009}
Mordasini, C., Alibert, Y., Benz W., \& Naef, D. 2009, \aap, 501, 1161

\bibitem[\protect\citeauthoryear{Mordasini et al.}{2012}]{Mordasinietal2012}
Mordasini, C., Alibert, Y., Georgy, C. et al. 2012, \aap, 547, A112

\bibitem[\protect\citeauthoryear{Ogihara et al.} {2013}]{Ogiharaetal2013} 
Ogihara, M., Inutsukar, S.-I., Kobayashi, H., 2013, \apj, 778, L9

\bibitem[\protect\citeauthoryear{Pepe et al.} {2002}]{pepe02} 
Pepe, F., Mayor, M., Galland, F., et al. 2002, \aap, 388, 632

\bibitem[\protect\citeauthoryear{Pepe et al.} {2013}]{Pepeetal2013} 
Pepe, F., Cameron, A. C., Latham, D. W. et al. 2013, Nature, 503, 377

\bibitem[\protect\citeauthoryear{Pollack et al.} {1996}]{Pollacketal1996} 
Pollack, J. B., Hubickyj, O., Bodenheimer, P., et al. 1996, Icarus, 124, 62

\bibitem[\protect\citeauthoryear{Rasio \& Ford}{1996}]{rasio96}
Rasio, F.A., \& Ford, E.B. 1996, Science, 274, 954

\bibitem[\protect\citeauthoryear{Rogers et al.}{2011}]{Rogersetal2011}
Rogers, L. A., Bodenheimer, P., Lissauer, J. J., \& Seager, S. 2011, \apj, 738, 59

\bibitem[\protect\citeauthoryear{Rowe et al.}{2014}]{Roweetal2014}
Rowe, J. F., Bryson, S. T., Marcy, G. W. et al. 2014, \apj, 784, 45

\bibitem[\protect\citeauthoryear{Raymond et al.}{2008}]{Raymondetal2008}
Raymond, S. N., Barnes, R., \& Mandell, A. M. 2008, \mnras, 384, 663

\bibitem[\protect\citeauthoryear{Santerne et al.} {2011}]{Santerneetal2011}
Santerne, A., D\'iaz, R. F., Bouchy, F. et al. 2011, \aap, 528, A63

%\bibitem[\protect\citeauthoryear{Sing} {2010}]{Sing2010}
%Sing, D. K. 2010, \aap, 510, A21

%Seageretal2007

\bibitem[Sanz-Forcada et al., 2011]{Sanz-Forcadaetal2011}
Sanz-Forcada, J., Micela, G., Ribas, I., et al. 2011, \aap, 531, A6

\bibitem[Seager et al., 2007]{Seageretal2007}
Seager, S., Kuchner, M., Hier-Majumder, C. A., \& Militzer, B., et al. 2007, \apj, 669, 1279

\bibitem[Sneden, 1973]{Sneden1973}
Sneden, C. A. 1973, Ph.D. Thesis, The University of Texas at Austin

\bibitem[Sozzetti et al., 2004]{Sozzettietal2004}
Sozzetti, A., Yong, D., Torres, G., et al. 2004, \apj, 616, L167

\bibitem[Sozzetti et al., 2006]{Sozzettietal2006}
Sozzetti, A., Yong, D., Carney, B. W., et al. 2006, \aj, 131, 2274

\bibitem[\protect\citeauthoryear{Sozzetti et al.} {2007}]{Sozzettietal2007}
Sozzetti, A., Torres, G., Charbonneau, D. et al. 2007, \apj, 664, 1190

\bibitem[\protect\citeauthoryear{Sousa et al.} {2010}]{Sousaetal2010}
Sousa, S. G., Alapini, A., Israelian, G., \& Santos, N. C. 2010, \aap, 512, A13

%\bibitem[\protect\citeauthoryear{Southworth} {2011}]{Southworth2011}
%Southworth, J. 2011, \mnras, 417, 2166

\bibitem[\protect\citeauthoryear{Southworth} {2012}]{Southworth2012}
Southworth, J. 2012, \mnras, 426, 1292

\bibitem[\protect\citeauthoryear{Steffen et al.} {2012}]{Steffenetal2012}
Steffen, J. H., Fabrycky, D. C., Ford, E. B., et al. 2012, \mnras,  421, 2342

\bibitem[\protect\citeauthoryear{Ter Braak} {2006}]{TerBraak2006}
Ter Braak, C. J. F. 2006, Statistics and Computing, 16, 239

\bibitem[\protect\citeauthoryear{Torres et al.} {2012}]{Torresetal2012}
Torres, G., Fischer, D. A., Sozzetti, A. et al. 2012, \apj, 757, 161

%\bibitem{ward97}
\bibitem[\protect\citeauthoryear{Ward}{1997}]{ward97}
Ward, W.R. 1997, Icarus, 126, 261

\bibitem[\protect\citeauthoryear{Zeng \& Sasselov}{2013}]{ZengSasselov2013}
Zeng, L., \& Sasselov, D. 2013, \pasp, 125, 227

}

\end{thebibliography}
\end{document}